%% file: covid-requirements20newsclips.tex
\documentclass[10pt]{article}

\usepackage[utf8]{inputenc}
\usepackage[T1]{fontenc}
\usepackage{url}
\usepackage{natbib}
\usepackage{xspace}
\usepackage{framed}
\usepackage{enumitem}

\usepackage{wojtek15logics,wojtek15other}

\usepackage{xcolor}

\newcommand{\cit}[1]{\cite[]{#1}}
\newcommand{\Covid}{COVID-19\xspace}
\newcommand{\WJ}[1]{{\color{red}#1}}
\renewcommand{\WJ}[1]{}
\newcommand{\MS}[1]{{\color{red}#1}}
\renewcommand{\MS}[1]{#1}
\newenvironment{myquote}{\begin{quote}\sf}{\end{quote}}

\title{A Survey of Requirements \\ for COVID-19 Mitigation Strategies \medskip \\
  Part I: Newspaper Clips}
\author{
  Wojciech Jamroga$^{1,2}$ \and David Mestel$^{1}$ \and Peter B. Roenne$^{1}$ \and Peter Y. A. Ryan$^{1}$ \and Marjan Skrobot$^{1}$
}
\date{
  ${}^1$ Interdisciplinary Centre on Security, Reliability and Trust, SnT, \\ University of Luxembourg
  \\
  ${}^2$ Institute of Computer Science, Polish Academy of Sciences, \\ Warsaw, Poland
}

\begin{document}

\maketitle

\tableofcontents

\section{Introduction}

The \Covid pandemic has influenced virtually all aspects of our lives.
Across the world, countries have applied wildly varying mitigation strategies for the epidemic, ranging from minimal intrusion in the hope of obtaining ``herd immunity'', to imposing severe lockdowns on the other extreme.
The strategies are based on various social, political, and technological instruments.
According to Digital Rights Tracker, as of June 2020:
\begin{itemize2}
\item Contact Tracing Apps were being used in 28 countries,
\item Alternative digital tracking measures were active in 35 countries,
\item Physical surveillance technologies were in use in 11 countries,
\item \Covid-related censorship had been imposed by 18 governments,
\item Internet shutdowns continued in 3 countries despite the outbreak
\cit{Top10VPN-Digital-Rights-2020-06-10}.
\end{itemize2}

It seems clear at the first glance what all those measures are trying to achieve, and what the criteria of success are.
But is it really that clear? Quoting an oft-repeated phrase, with \Covid we fight
\begin{myquote}
an unprecedented threat to \emph{health} and \emph{economic stability}\footnote{
  The highlighting of selected words and phrases is due to the authors of this report throughout the text. }
\cit{Brookings-Tech-Stream-CTAs-2020-04-27}
\end{myquote}
While fighting it, we must
\begin{myquote}
Protect \emph{privacy}, \emph{equality} and \emph{fairness} 
\cit{Nature-Comment-Ethical-guidelines-2020-06-04}
\end{myquote}
and
\begin{myquote}
do a \emph{coordinated assessment} of \emph{usefulness}, \emph{effectiveness}, \emph{technological readiness}, \emph{cyber security risks} and threats to \emph{fundamental freedoms} and \emph{human rights}
\cit{euobserver-Dutch-soap-opera-2020-05-07}
\end{myquote}
\MS{Moreover, we should do it in a way that
\begin{myquote}
needs to be \emph{ethical}, \emph{trustworthy}, \emph{locally rooted}, and \emph{adaptive to new data on what works}.
\cit{Why-people-dont-2020-11-12}
\end{myquote}}
Taken together, this is hardly a straightforward set of goals and requirements.
Thus, paraphrasing \cit{euobserver-Dutch-soap-opera-2020-05-07}, one may ask:
\begin{quote}
\textbf{What problem does a COVID mitigation strategy solve exactly?}
\end{quote}

Even a quick survey of news articles, manifestos, and research papers published since the beginning of the pandemic reveals a diverse landscape of postulates and opinions. Some authors focus on medical goals, some on technological requirements; others are more concerned by the economic, social, or political impact of a containment strategy.
The actual stance is often related to the background of the author (in case of a researcher) or their information sources (in case of a journalist).
We postulate that one should {identify the relevant requirements} before committing to a particular mitigation strategy.
In fact, this should be done before any informal or formal analysis and assessment of the available strategies, with as little bias as possible.

One way to achieve it is through an overview of what is considered relevant by the general public, and referred to in the media.
To this end, we have collected a number of quotes and ``news clips'' touching on the topic.
We looked especially at the most popular IT solution proposed to contain the epidemic, namely \emph{digital contact tracing apps}.
The idea has generated a fair amount of controversy, and for a while virtually dominated the discussions and press coverage of \Covid containment strategies.
Eventually contact tracing apps have been deployed on a large scale, with several hundred million downloads across the world,\footnote{
  The Indian app Aarogya Setu alone has already had more that 100 million downloads~\cit{TimesofIndia-100million-2020-05-13}. }
leading to the discovery of various flaws and even more discussion on what characterizes acceptable and effective containment measures.

In this paper, we present some snippets from the press and Internet media, published between March and June 2020, that mention the possible goals and requirements for a mitigation strategy.
The snippets are sorted thematically into several categories, such as health-related goals, social and political impact, civil rights, ethical requirements, etc.
While most of the presented news clips focus on digital contact tracing for \Covid, it is easy to see that the relevance of the requirements goes beyond contact tracing apps, and applies to any large scale epidemic: also the ones that may happen in the future.

We emphasize that \textbf{we do not endorse the opinions being presented in the quotes}.
We also \textbf{do not comment on their content}.
We merely use the quotes to collect relevant keywords and conceptual categories.
The assessment of how the requirements are addressed by different mitigation strategies, and which strategy to select, is a matter for systematic analysis that should follow in the next step.


\input{epid+imed+psy}


\input{econ+soc}


\input{eth+civ+leg}


\input{priv+data}


\input{user+tech}

\input{eval+know}


\section{Conclusions}

In this report, we make the first step towards a systematic analysis of strategies for effective and trustworthy mitigation of the current pandemic.
The strategies may incorporate medical, social, economic, as well as technological measures. Consequently, there is a large number of medical, social, economic, and technological requirements that must be taken into account when deciding which strategy to adopt.
Being computer scientists who specialize in information security, we find the latter kind of requirements most obvious, especially with respect privacy and security of information flow.
This is exactly the pitfall that computer scientists must avoid.
It is essential to realize that the goals (and acceptability criteria) for a mitigation strategy are much more diverse, and consciously choose a solution that satisfies the multiple criteria to a reasonable degree.
\WJ{Mention multicriterial optimization and Pareto frontier?}

In a forthcoming companion paper\WJ{\cit{CITE}}, we will present a digest of the requirements, derived from the news clips, and a preliminary take on their formal specification.

\para{Acknowledgments.}
The authors acknowledge the support of the Luxembourg National Research Fund (FNR) under the COVID-19 project SmartExit,
and the support of the National Centre for Research and Development Poland (NCBR) and the Luxembourg National Research Fund (FNR), under the PolLux/FNR-CORE project STV (POLLUX-VII/1/2019).
\WJ{Ack Ala for helping with bibentries?}

\WJ{Underrepresented categories:
- protecting health of individuals
- wellbeing, mental health, social relationships
- economic stability (to a lesser extent)
}

\WJ{In the companion paper -- explain methodology: looking for relevant phrases that appear in ordinary texts; no particular methodology of source selection}

\nocite{BBC-Coronavirus-privacy-2020-03-05}
\nocite{POLITICO-Polands-2020-04-02}
\nocite{Brookings-Tech-Stream-CTAs-2020-04-27}
\nocite{Telecoms-Unlike-France-2020-04-27}
\nocite{Telecoms-UK-2020-04-28}
\nocite{helsenorge-Together-2020-04-28}
\nocite{Tabletowo-Aplikacja-2020-04-29}
\nocite{WashingtonPost-Most-Americans-2020-04-29}
\nocite{MatrixChambers-Legal-Advice-2020-05-03}
\nocite{NCSC-nhs-explainer-2020-05-04}
\nocite{Register-UK-finds-2020-05-05}
\nocite{Cybernetica-Proposes-2020-05-06}
\nocite{Panoptykon-ProteGo-Safe-2020-05-06}
\nocite{SDZ-Corona-App-2020-05-06}
\nocite{euobserver-Dutch-soap-opera-2020-05-07}
\nocite{MIT-covid-tracing-2020-05-07}
\nocite{Wired-Just-how-2020-05-12}
\nocite{Gizmodo-UKs-Contact-Tracing-2020-05-13}
\nocite{BBC-News-Why-Indias-2020-05-15}
\nocite{POLITICO-States-struggle-2020-05-17}
\nocite{Nature-Can-they-slow-2020-05-19}
\nocite{Guardian-Covidsafe-2020-05-23}
\nocite{Nature-Comment-Ethical-guidelines-2020-06-04}
\nocite{RTL-LU-Lockdowns-averted-2020-06-09}
\nocite{Politico-Privacy-fears-2020-06-04}
\nocite{MIT-Technology-coronavirus-apps-2020-06-05}
\nocite{Top10VPN-Digital-Rights-2020-06-10}
\nocite{LeMonde-StopCovid-2020-06-10}
\nocite{Politico-Google-Apple-2020-06-10}
\nocite{NatureComment-Ten-reasons-2020-05-21}
\nocite{AdaLovelace-Something-to-declare-2020-06-02}
\nocite{EDRi-COVID-Tech-2020-06-10}
\nocite{Guardian-Norway-2020-06-15}
\nocite{BBCNews-Alarm-Kuwait-2020-06-16}
\nocite{EuropeanCommission-interoperability-2020-06-16}
\nocite{Guardian-democracies-2020-06-16}
\nocite{ScientificAmerican-AntibodyTests-2020-06-17}
\nocite{Guardian-UK-abandons-2020-06-18}
\nocite{RTL-German-app-2020-06-25}

\bibliographystyle{plainnat}
\bibliography{covid}

\end{document}

%% file: epid+imed+psy.tex

\section{Epidemiological and Health-Related Concerns}
\label{sec:epid}

Clearly, an epidemic is first and foremost a threat to people's health and lives. Accordingly, we begin with media clips related to this aspect of \Covid mitigation strategies.

\subsection{Epidemiological Goals}

Containment measures (and contact tracing apps in particular) should slow the spread of the virus, decrease the transmission rate, and save lives:
\begin{myquote}
Since the outbreak of COVID-19, governments around the world have implemented a range of digital tracking, physical surveillance and censorship measures in a bid to \emph{slow the spread of the virus}.
\cit{Top10VPN-Digital-Rights-2020-06-10}
\end{myquote}

\begin{myquote}
Lockdowns \emph{prevented} around 3.1 million \emph{deaths} in 11 European countries, according to a new modelling study
\cit{RTL-LU-Lockdowns-averted-2020-06-09}
\end{myquote}

\begin{myquote}
Researchers also calculated that the interventions had \emph{caused the reproduction number} -- how many people someone with the virus infects -- \emph{to drop} by an average of 82 percent, to below 1.0.
\cit{RTL-LU-Lockdowns-averted-2020-06-09}
\end{myquote}

\begin{myquote}
Contact tracing can be an important component of an \emph{epidemic response} especially when the prevalence of infection is low. Such efforts are most \emph{effective} where testing is rapid and widely available and when infections are relatively rare.
\cit{Brookings-Tech-Stream-CTAs-2020-04-27}
\end{myquote}

\begin{myquote}
To \emph{best meet public health needs}, digital technology should be able to \emph{trace the spread of the virus}, \emph{identify dangerous Covid-19 clusters} and \emph{limit further transmission}. The essential goal is to \emph{register contacts between potential carriers and those who might be infected}.
\cit{Guardian-democracies-2020-06-16}
\end{myquote}

\begin{myquote}
designed and built by the NHS to help \emph{slow the spread of the coronavirus}
\cit{NCSC-nhs-explainer-2020-05-04}
\end{myquote}

\begin{myquote}
you’ll be helping to \emph{slow the transmission of the coronavirus}.
\cit{NCSC-nhs-explainer-2020-05-04}
\end{myquote}

\MS{%
\begin{myquote}
even finding a fraction of cases through contact tracing will help \emph{slow the virus’s spread}.
\cit{Contact-Tracing-Is-Failing-2020-10-05}
\end{myquote}
}

\begin{myquote}
Contact tracing via smartphone is a powerful way to \emph{tackle the spread of coronavirus}, but it mustn’t be done at the expense of individual civil rights.
\cit{Telecoms-Unlike-France-2020-04-27}
\end{myquote}

\begin{myquote}
Smittestopp is an app that will \emph{help the health authorities to limit the transmission of coronavirus} and alert users with text messages about close contact with infected persons.
\cit{helsenorge-Together-2020-04-28}
\end{myquote}

\subsection{Effectiveness of Epidemic Response}

The measures should be effective in containing the epidemic:

\begin{myquote}
what's missing in most debates is how \emph{effective} the apps actually are
\cit{euobserver-Dutch-soap-opera-2020-05-07}
\end{myquote}

\begin{myquote}
We and many others have pointed out a host of pitfalls for voluntary, self-reported coronavirus apps of the kind Apple, Google, and others contemplate. First, app notifications of contact with COVID-19 are likely to be simultaneously both \emph{over-} and \emph{under-inclusive}. (…) Individuals may be flagged as having contacted one another despite very low possibility of transmission—such as when the individuals are separated by walls porous enough for a Bluetooth signal to penetrate. Nor do the systems account for when individuals take precautions, such as the use of personal protective equipment, in their interactions with others. At least as problematic is the issue of \emph{false negatives} (…) Contact-tracing apps therefore \emph{cannot offer assurance} that going out is safe, just because no disease has been reported in the vicinity.
\cit{Brookings-Tech-Stream-CTAs-2020-04-27}
\end{myquote}

\begin{myquote}
Inaccurate results are likely to lead to a high number of false positives and negatives that may \emph{adversely impact the relaxation of lockdown measures}.
\cit{Top10VPN-Digital-Rights-2020-06-10}
\end{myquote}

\begin{myquote}
On the Isle of Wight, the NHS contact tracing app \emph{is making a difference}. Around 25 people per day are being tested for coronavirus after reporting it through their phones
\cit{Wired-Just-how-2020-05-12}
\end{myquote}

\MS{%
\begin{myquote}
Initially, the hope was that contact tracing apps would be \emph{a silver bullet} for contact tracing. Evidence shows this is not the case. Many European countries have been using these apps for months with \emph{limited success}.
\cit{How-to-Fix-COVID-2020-12-07}
\end{myquote}
}

\MS{%
\begin{myquote}
Contact tracing apps have struggled with \emph{low adoption rates}, \emph{issues with the accuracy of the Bluetooth technology} on which the apps rely to detect when app users come into contact with each other, and ensuring that app users remember to bring their phones with them and to upload their coronavirus test results. \cit{How-to-Fix-COVID-2020-12-07}
\end{myquote}
}

\MS{%
\begin{myquote}
Tracking those exposed is so far behind the virus raging in most places that many public health officials believe the money and personnel involved would be better spent on other resources, like increasing test sites, helping schools prepare for reopening and educating the public about mask wearing. 
\cit{Contact-Tracing-Is-Failing-2020-10-05}
\end{myquote}
}

This connects to \emph{timeliness} which is an essential feature of an effective response strategy:

\MS{%
\begin{myquote}
It’s a \emph{race against time}
\cit{Contact-Tracing-Is-Failing-2020-10-05}
\end{myquote}
}

\subsection{Information Flow to Counter the Epidemic}

The strategy should support rapid information flow to the most concerned:
\begin{myquote}
As the global fight against the COVID-19 pandemic continues, much of the world is pinning its hopes of {easing lockdowns} on being able to \emph{quickly identify people who might have been exposed to the virus}.
\cit{Nature-Can-they-slow-2020-05-19}
\end{myquote}

\begin{myquote}
Technologies to \emph{rapidly alert people} when they have been in contact with someone carrying the coronavirus SARS-CoV-2 are part of a strategy to \emph{bring the pandemic under control}.
\cit{Nature-Comment-Ethical-guidelines-2020-06-04}
\end{myquote}

\begin{myquote}
some public health and consumer advocacy groups have urged the use of apps that allow people who test positive for Covid-19 anonymously message recent contacts.
\cit{POLITICO-States-struggle-2020-05-17}
\end{myquote}

\begin{myquote}
a retired home-care nurse (…) said she would welcome an app to \emph{help spread word of possible exposure}.
\cit{WashingtonPost-Most-Americans-2020-04-29}
\end{myquote}

\begin{myquote}
Smittestopp is an app that will {help the health authorities} to {limit the transmission of coronavirus} and \emph{alert users with text messages about close contact with infected persons}.
\cit{helsenorge-Together-2020-04-28}
\end{myquote}

\subsection{Tradeoffs}

Clearly, there are tradeoffs between epidemiological efficiency and other concerns, as we will also see in the subsequent sections:

\begin{myquote}
The immediate goal for governments and tech companies is to \emph{strike the right balance} between \emph{privacy} and the \emph{effectiveness of an application to limit the spread of Covid-19}.
\cit{Guardian-democracies-2020-06-16}
\end{myquote}

\begin{myquote}
So far the clear evidence is that greater control of populations has \emph{worked better at stopping the coronavirus spread} than a more relaxed attitude
\cit{Register-UK-finds-2020-05-05}
\end{myquote}

\begin{myquote}
Some states are raising concerns that the [Google and Apple] app won’t allow them to \emph{collect enough information} due to privacy concerns.
\cit{POLITICO-States-struggle-2020-05-17}
\end{myquote}

\begin{myquote}
China's Health Code system (...) {records a user's spending history} in order to \emph{deter them from breaking quarantine}
\cit{BBC-News-Why-Indias-2020-05-15}
\end{myquote}

\subsection{Monitoring Pandemic and Containment Strategy}

The containment strategy should enable monitoring the state of the pandemic, the behavior of people, and the effectiveness of the strategy itself:

\begin{myquote}
Most states are waiting to see whether the Bluetooth-based release from Apple and Google, which is supposed to automatically notify people when they come close to someone who’s tested positive, will be an effective way to \emph{monitor outbreaks}.
\cit{POLITICO-States-struggle-2020-05-17}
\end{myquote}

\begin{myquote}
These serology tests can provide important data on \emph{how COVID-19 is spreading through a population}. There is also hope that the presence of certain antibodies may signify immunity to future infection
\cit{ScientificAmerican-AntibodyTests-2020-06-17}
\end{myquote}

\begin{myquote}
"{The police won’t have to monitor the places of quarantine}," said Mateusz Morawiecki, the Polish prime minister, when announcing the country's coronavirus app would become mandatory. \emph{"We will know if people are following the rules."}
\cit{POLITICO-Polands-2020-04-02}
\end{myquote}

\noindent
Effective monitoring is often at odds with other requirements, especially privacy:
\begin{myquote}
There are of course advantages to [the centralized] approach [of contact tracing], \emph{models can be adapted quicker and additional analysis can be performed}, but the question which remains is whether this outweighs the \emph{risk to security and privacy}
\cit{Telecoms-UK-2020-04-28}
\end{myquote}

\subsection{Specific Goals of Digital Contact Tracing}

Digital measures for \Covid mitigation often have more specific goals and desirable properties.

\begin{myquote}
Ultimately, contact tracing is a \emph{public health intervention}, not an {individual health} one. It can \emph{reduce the spread of disease through the population}
\cit{Brookings-Tech-Stream-CTAs-2020-04-27}
\end{myquote}

\begin{myquote}
[Contact tracing apps are] a way to \emph{enhance traditional forms of contact tracing} to \emph{find potential new infections}
\cit{WashingtonPost-Most-Americans-2020-04-29}
\end{myquote}

\MS{%
\begin{myquote}
The goal of contact tracing for Covid-19 is to \emph{reach people} who have spent more than 15 minutes within six feet of an infected person and \emph{ask them to quarantine} at home voluntarily for two weeks even if they test negative, monitoring themselves for symptoms during that time.
\cit{Contact-Tracing-Is-Failing-2020-10-05}
\end{myquote}

\begin{myquote}
Traditional contact tracing works by interviewing those infected with COVID-19 about whom they have encountered during the prior two weeks. The contact tracers then contact those who have been exposed to \emph{notify them of exposure}. \cit{How-to-Fix-COVID-2020-12-07}
\end{myquote}
}

\MS{%
\subsection{Prerequisites for Successful Contact Tracing}

Contact tracing at scale works successfully only under certain conditions:

\begin{myquote}
Contact tracing generally works best, public health experts say, when a disease is \emph{easily detected from its onset}. That is often impossible with the coronavirus because a large percentage of those infected have no symptoms.
\cit{Contact-Tracing-Is-Failing-2020-10-05}
\end{myquote}

\begin{myquote}
Perhaps most harmful to the effort have been the persistent \emph{delays} in getting the results of diagnostic tests. Often by the time an individual tests positive, it’s too late for the health care workers tracking that person to do anything. In contrast, when sports teams and staff of the White House test people constantly, with fast turnarounds, contact tracing is \emph{instant} and \emph{effective}.
\cit{Contact-Tracing-Is-Failing-2020-10-05}
\end{myquote}

\begin{myquote}
Contact tracing, a cornerstone of the public health arsenal to tamp down the coronavirus across the world, \emph{has largely failed} in the United States; the \emph{virus’s pervasiveness} and major \emph{lags in testing} have rendered the system almost pointless. In some regions, large swaths of the population have \emph{refused to participate} or cannot even be \emph{located}, further hampering health care workers.
\cit{Contact-Tracing-Is-Failing-2020-10-05}
\end{myquote}

\begin{myquote}
Some public health experts now believe that, at the very least, testing and contact tracing need to be \emph{scaled back in places with major outbreaks}. In some places, they say the effort may never succeed.
\cit{Contact-Tracing-Is-Failing-2020-10-05}
\end{myquote}
}


%% file: econ+soc.tex

\section{Economic and Social Impact}
\label{sec:econ+soc}

Most measures to contain the epidemic are predominantly social (lockdown being a prime example), and have strong social and economic impact.

\subsection{Economic Stability}

The impact of the containment strategy on the economy is of prime importance:
\begin{myquote}
The unprecedented threat from the novel coronavirus has confined many Americans to their homes, distancing them from one another at \emph{great cost to local economies} and {personal well-being}.
\cit{Brookings-Tech-Stream-CTAs-2020-04-27}
\end{myquote}

\begin{myquote}
In a separate study, also published in Nature, researchers from UC Berkeley used a different method -- econometric modelling used to assess how policies \emph{affect economic growth} -- to evaluate containment policies in China, South Korea, Italy, Iran, France and the United States.
\cit{RTL-LU-Lockdowns-averted-2020-06-09}
\end{myquote}

\begin{myquote}
[Contact tracing apps] help make \emph{resumption of economic and social activities safer} in the months ahead.
\cit{WashingtonPost-Most-Americans-2020-04-29}
\end{myquote}

\begin{myquote}
[Australian Prime Minister, Scott Morrison:] “If you want to \emph{return to a more liberated economy and society}, it is important that we get increased numbers of downloads when it comes to the Covidsafe app.”
\cit{Guardian-Covidsafe-2020-05-23}
\end{myquote}

\subsection{Impact on Society}

Regarding the available measures in general, and contact tracing apps in particular:
\begin{myquote}
we know very little about them or how they could \emph{affect society}.
\cit{MIT-covid-tracing-2020-05-07}
\end{myquote}

\begin{myquote}
As the global fight against the COVID-19 pandemic continues, much of the world is pinning its hopes of \emph{easing lockdowns} on being able to quickly identify people who might have been exposed to the virus.
\cit{Nature-Can-they-slow-2020-05-19}
\end{myquote}

\begin{myquote}
[The Covidsafe app] was sold as the key to \emph{unlocking restrictions} (…) but as the country begins to open up, the role of the Covidsafe app in the recovery seems to have dropped to marginal at best.
\cit{Guardian-Covidsafe-2020-05-23}
\end{myquote}

\begin{myquote}
The health minister, Greg Hunt, tweeted that [the Covidsafe app] was the key to \emph{being allowed to go back to watching football}.
\cit{Guardian-Covidsafe-2020-05-23}
\end{myquote}

\subsubsection{Disinformation and Information Abuse}

\begin{myquote}
We worry that contact-tracing apps will serve as vehicles for \emph{abuse and disinformation}, while providing a \emph{false sense of security} to justify reopening local and national economies well before it is safe to do so.
\cit{Brookings-Tech-Stream-CTAs-2020-04-27}
\end{myquote}

\begin{myquote}
Since the outbreak of COVID-19 there has been a rapid acceleration in the \emph{spread of mis- and disinformation}. To \emph{control this}, governments and social media platforms have sought to stringently \emph{regulate online content} and \emph{promote official facts and figures from international health organisations}.
\cit{Top10VPN-Digital-Rights-2020-06-10}
\end{myquote}

\subsubsection{Potential for Discrimination and Creation of Social Divides}

\begin{myquote}
There is also a very real danger that these voluntary surveillance technologies will effectively become \emph{compulsory for any public and social engagement}.
\cit{Brookings-Tech-Stream-CTAs-2020-04-27}
\end{myquote}

\begin{myquote}
there is a real risk that these mobile-based apps can \emph{turn unaffected individuals into social pariahs, restricted from accessing public and private spaces or participating in social and economic activities}.
\cit{Brookings-Tech-Stream-CTAs-2020-04-27}
\end{myquote}

\begin{myquote}
\emph{protecting} those \emph{communities who can be (…) harmed by the collection and exploitation of personal data}.
\cit{Brookings-Tech-Stream-CTAs-2020-04-27}
\end{myquote}

\begin{myquote}
Protections need to be put in place to expressly \emph{prohibit economic and social discrimination on the basis of information and technology designed to address the pandemic}.
\cit{Brookings-Tech-Stream-CTAs-2020-04-27}
\end{myquote}

\subsection{Political Impact}

The pandemic creates new space for political manipulation and changes the distribution of power:
\begin{myquote}
the issue of malicious use is paramount—particularly given this current climate of disinformation, astroturfing, and \emph{political manipulation}. Imagine an unscrupulous political operative who wanted to dampen voting participation in a given district, or a desperate business owner who wanted to stifle competition. Either could falsely report incidences of coronavirus without much fear of repercussion. Trolls could sow chaos for the malicious pleasure of it. Protesters could trigger panic as a form of civil disobedience. A foreign intelligence operation could shut down an entire city by falsely reporting COVID-19 infections in every neighborhood.
\cit{Brookings-Tech-Stream-CTAs-2020-04-27}
\end{myquote}

\begin{myquote}
In the long run, however, this poses a far more fundamental question: \emph{how much can the decisions of sovereign democratic countries be overruled by technology companies} [i.e., Google and Apple]?
\cit{Guardian-democracies-2020-06-16}
\end{myquote}

\subsection{Wellbeing, Mental Health, and Social Relationships}

\begin{myquote}
The unprecedented threat from the novel coronavirus has \emph{confined many Americans to their homes}, \emph{distancing them from one another} at {great cost to local economies} and \emph{personal well-being}.
\cit{Brookings-Tech-Stream-CTAs-2020-04-27}
\end{myquote}

\begin{myquote}
we all long for \emph{freedom from lockdowns and home confinement}
\cit{euobserver-Dutch-soap-opera-2020-05-07}
\end{myquote}

\subsection{Costs, Human Resources, Logistics}

The financial costs incurred by the containment measures, as well as the engaged human resources, are another important factor.
\begin{myquote}
The government has been forced to abandon a centralised coronavirus contact-tracing app after \emph{spending three months and millions of pounds} on technology that experts had repeatedly warned would not work. (…) [T]he health secretary said the government would not \emph{“put a date”} on when the new app may be launched, although officials conceded it was likely to be in the autumn or winter.
\cit{Guardian-UK-abandons-2020-06-18}
\end{myquote}

\begin{myquote}
"\emph{The police won’t have to monitor the places of quarantine}," said Mateusz Morawiecki, the Polish prime minister, when announcing the country's coronavirus app would become mandatory. "We will know if people are following the rules."
\cit{POLITICO-Polands-2020-04-02}
\end{myquote}

\begin{myquote}
The lure of \emph{automating the painstaking process} of [manual] contact tracing is apparent.
\cit{Brookings-Tech-Stream-CTAs-2020-04-27}
\end{myquote}

\MS{%
\begin{myquote}
In Michigan, Ottawa County's public health department \emph{automated} {contact tracing symptom checks and follow-ups}. \cit{How-to-Fix-COVID-2020-12-07}
\end{myquote}

\begin{myquote}
As coronavirus cases reach new peaks, surpassing 200,000 cases a day in the United States, \emph{public health departments are overwhelmed}. Departments are racing to hire yet more contact tracers and some are even asking \emph{people to do their own contact tracing and notification}. \cit{How-to-Fix-COVID-2020-12-07}
\end{myquote}
}

The logistics of the strategy (in particular, coordination and cooperation between different institutions and authorities) also plays a role:
\begin{myquote}
U.S. efforts (…) are running into an all-too-familiar problem that has plagued the pandemic response: a lack of \emph{national coordination}.
\cit{Politico-Google-Apple-2020-06-10}
\end{myquote}

\begin{myquote}
the federal government has so far failed to institute concrete \emph{privacy standards}.
\cit{Politico-Google-Apple-2020-06-10}
\end{myquote}

Similarly, the European Commission postulated
\begin{myquote}
a set of technical specifications to ensure a \emph{safe exchange of information between national contact tracing apps} based on a decentralised architecture [to] \emph{work seamlessly when users travel to another EU country} which also follows the decentralised approach.
\cit{EuropeanCommission-interoperability-2020-06-16}
\end{myquote}

%% file: eth+civ+leg.tex

\section{Compliance with Norms, Rights, and Obligations}

In this section, we look at requirements that aim at the long-term robustness and resilience of the social structure, such as ethical and legal requirements.

\subsection{Ethical Aspects}
\label{sec:ethic}

The mitigation strategy must be \emph{ethically justifiable}~\cit{Nature-Comment-Ethical-guidelines-2020-06-04}. Importantly:
\begin{myquote}
\emph{ethical and social considerations} [other than privacy] must not be cast aside in the rush to quell the pandemic. For instance, contact-tracing apps should be \emph{available and accessible to anyone}, irrespective of the technology needed or their level of digital literacy.

(...)
Even in a crisis, a ‘try-everything’ approach is dangerous when it ignores the real costs, including \emph{serious and long-lasting harms to fundamental rights and freedoms}, and the opportunity costs of \emph{not devoting resources to something else}.
\cit{Nature-Comment-Ethical-guidelines-2020-06-04}
\end{myquote}

\begin{myquote}
Some of these [mitigation measures] may well be \emph{proportionate}, \emph{necessary} and \emph{legitimate} during these unprecedented times. However, others have been rushed through legislative bodies and implemented \emph{without adequate scrutiny}.
\cit{Top10VPN-Digital-Rights-2020-06-10}
\end{myquote}

\begin{myquote}
the data agency Datatilsynet (…)  said the restricted spread of coronavirus in Norway, as well as the app’s limited effectiveness due to the small number of people using it, meant the invasion of privacy resulting from its use was \emph{disproportionate}.
\cit{Guardian-Norway-2020-06-15}
\end{myquote}

\noindent
Note that the ethical considerations do not necessarily oppose strict measures of containing the epidemic. In particular:
\begin{myquote}
Temporarily restricting some fundamental rights and freedoms \emph{might be ethically justifiable} in the context of hastening the end of the pandemic. Quarantining individuals, for example, helps to prevent the spread of the disease. Arguably, it might be unethical not to use digital tracing apps when necessary. Nevertheless, much depends on the effectiveness of the app, the goal pursued, the type of system and the context in which it will be deployed.

(...) To be ethical, a contact-tracing app must abide by four principles: it must be \emph{necessary}, \emph{proportional}, \emph{scientifically valid} and \emph{time-bound}. These principles are derived from the European Convention on Human Rights, the International Covenant on Civil and Political Rights (ICCPR) and the United Nations Siracusa Principles, which specify the provisions in the ICCPR that limit how it can be applied.
\cit{Nature-Comment-Ethical-guidelines-2020-06-04}
\end{myquote}

\begin{myquote}
An app’s \emph{implementation strategy} and \emph{impact} must also be considered. Something that looked good on paper can turn out to be ineffective in practice. (…) \emph{If an app fails, it becomes unnecessary, and thus unethical}.
\cit{Nature-Comment-Ethical-guidelines-2020-06-04}
\end{myquote}

Moreover, the terms and conditions of such an app should be \emph{just}~\cit{POLITICO-Polands-2020-04-02}.
One must also keep in mind that ethical norms and their interpretation vary depending on the actual place, time, and social context.
\begin{myquote}
In practice, there will be \emph{trade-offs}. These will \emph{depend on the laws, values, attitudes and norms in different regions}, as well as on \emph{changes over time} in the spread and scale of the virus and the available technology. (…) Similarly, what was ethically justifiable in one place yesterday might not be so tomorrow.
\cit{Nature-Comment-Ethical-guidelines-2020-06-04}
\end{myquote}

\subsubsection{Detailed Criteria}

The detailed guidelines for ethical justifiability of a contact-tracing app from~\cit{Nature-Comment-Ethical-guidelines-2020-06-04} are presented below.

\input{ethical-guidelines}

\subsection{Civil Rights and Fundamental Freedoms}

An important aspect of ethical requirements is to protect rights and liberties:
\begin{myquote}
Contact tracing via smartphone is a powerful way to tackle the spread of coronavirus, but it \emph{mustn’t be done at the expense of individual civil rights}.
\cit{Telecoms-Unlike-France-2020-04-27}
\end{myquote}

\begin{myquote}
Is [the app] \emph{mandatory}? How \emph{private} is the app? Are citizens’ rights being \emph{safeguarded}?
\cit{MIT-covid-tracing-2020-05-07}
\end{myquote}

\noindent
According to a legal opinion on smartphone contact tracing and other data driven proposals:
\begin{myquote}
A de-centralised smartphone contact tracing system – the type contemplated by “DP-3T” and being considered by governments across Europe and also Apple and Google – would be likely to \emph{comply with both human rights and data protection laws}. In contrast, a centralised smartphone system – which is the current UK Government proposal – is a \emph{greater interference with fundamental rights} and would require \emph{significantly greater justification to be lawful}. That justification has not yet been forthcoming.
\cit{MatrixChambers-Legal-Advice-2020-05-03}
\end{myquote}

\begin{myquote}
Any attempt to introduce ‘immunity passports’ would be a \emph{dramatic measure, both socially and legally}. It would need a \emph{clear scientific basis} and would also have to \emph{address} the significant \emph{impact on fundamental rights} including the \emph{risk of indirect discrimination}.
\cit{MatrixChambers-Legal-Advice-2020-05-03}
\end{myquote}

\noindent
For instance, the designers of the NCSC app developed in the UK declared that:
\begin{myquote}
[installing the app is] entirely \emph{voluntary}
\cit{NCSC-nhs-explainer-2020-05-04}
\end{myquote}
and
\begin{myquote}
You can \emph{decide if you want to tell the app} that you’re suffering from coronavirus symptoms.
\cit{NCSC-nhs-explainer-2020-05-04}
\end{myquote}

\noindent
The reality, however, is often far removed from the intended properties, as the same NCSC app illustrates:
\begin{myquote}
as soon as someone agrees to share their information with UK government – by claiming to feel unwell and hitting a big green button – 28 days of data from the app is given to a central server from where \emph{it can never be recovered}. That data, featuring all the unique IDs you've encountered in that period and when and how far apart you were, \emph{becomes the property of NCSC} (…) [The NCSC chief exec Matthew Gould] also admitted that \emph{the data will not be deleted}, UK citizens \emph{will not have the right to demand it is deleted}, and it can or will be \emph{used for "research" in future}.
\cit{Register-UK-finds-2020-05-05}
\end{myquote}

\subsection{Threats to Fundamental Freedoms}

\begin{myquote}
In an attempt to slow the spread of COVID-19, governments around the world are also adopting \emph{increasingly extensive physical surveillance measures}. These include the deployment of facial recognition cameras equipped with heat sensors, surveillance drones used to monitor citizens’ movements, and extensive CCTV networks in a bid to help enforce curfews.
\cit{Top10VPN-Digital-Rights-2020-06-10}
\end{myquote}

\begin{myquote}
India has made it \emph{mandatory} for government and private sector employees to download [its contact-tracing app].
\cit{BBC-News-Why-Indias-2020-05-15}
\end{myquote}

\begin{myquote}
Using a phone's Bluetooth and location data, Aarogya Setu lets users know if they have been near a person with Covid-19 by scanning a database of known cases of infection. The data is then \emph{shared with the government}.
\cit{BBC-News-Why-Indias-2020-05-15}
\end{myquote}

\begin{myquote}
To register, users \emph{have to give their name, gender, travel history, telephone number and location}.
(...)
While your name and number won't be made public, the app does \emph{collect this information}, as well as \emph{your gender, travel history and whether you're a smoker}.
\cit{BBC-News-Why-Indias-2020-05-15}
\end{myquote}

\begin{myquote}
China's Health Code system, which \emph{records a user's spending history} in order to deter them from breaking quarantine, is \emph{invasive}.
\cit{BBC-News-Why-Indias-2020-05-15}
\end{myquote}

\begin{myquote}
Many regions that have made contact tracing a key part of their COVID-19 playbook — including China, South Korea, Taiwan and Israel — have \emph{empowered contact tracers with sensitive details of infected people}, including CCTV footage, credit-card transactions and location data from mobile-phone carriers.
\cit{Nature-Can-they-slow-2020-05-19}
\end{myquote}

\begin{myquote}
several governments have also co-opted the rise of mis/disinformation to justifying \emph{censorship practices which seek to silence critics and control the flow of information}.
\cit{Top10VPN-Digital-Rights-2020-06-10}
\end{myquote}

\subsection{Legal Requirements}

\begin{myquote}
UK finds itself almost alone with centralized virus contact-tracing app that (...) may be \emph{illegal}
\cit{Register-UK-finds-2020-05-05}
\end{myquote}

\begin{myquote}
The UK Government’s announcements in March and April for sharing health data between the private and public sector appear to be \emph{flawed}. This means such data sharing is potentially not \emph{in compliance with legal requirements}. Further information needs to be provided to ensure compliance and a \emph{data impact assessment should be conducted and made public}.
\cit{MatrixChambers-Legal-Advice-2020-05-03}
\end{myquote}

\begin{myquote}
the UK’s own contact-tracing app \emph{runs roughshod over regional [GDPR] regulations} and could, hypothetically, \emph{lead to a sweeping, NHS-wide-fine} in its current state.
\cit{Gizmodo-UKs-Contact-Tracing-2020-05-13}
\end{myquote}

%% file: ethical-guidelines.tex
\begin{framed}\sf\small
\noindent
\textbf{Principles: is this the right app to develop?}

\begin{enumerate}
\item Is it \emph{necessary}?

• Yes, it must be developed to save lives (+).

• No, there are better solutions (–).

\item Is it \emph{proportionate}?

• Yes, the gravity of the situation justifies the potential negative impact (+).

• No, the potential negative impact is disproportionate to the situation (–).

\item Is it \emph{sufficiently effective}, \emph{timely}, \emph{popular} and \emph{accurate}?

• Yes, evidence shows that it will work, is timely, will be adopted by enough people and yields accurate data and insights (+).

• No, it does not work well, is available too late or too early, will not be used widely, and is likely to collect data that have false positives and/or false negatives (–).

\item Is it \emph{temporary}?

• Yes, there is an explicit and reasonable date on which it will cease (+).

• No, it has no defined end date (–).
\end{enumerate}

\noindent
\smallskip
\textbf{Requirements: is this app being developed in the right way?}

\begin{enumerate}[start=5]
\item Is it \emph{voluntary}?

• Yes, it is optional to download and install (+).

• No, it is mandatory and people can be penalized for non-compliance (–).

\item Does it \emph{require consent}?

• Yes, people have complete choice over what data are shared and when, and can change this at any time (+).

• No, default settings are to share everything all the time, and this cannot be altered (–).

\item Are the \emph{data kept private} and \emph{users’ anonymity preserved}?

• Yes, data are anonymous and held only on the user’s phone. Others who have been in contact are notified only that there is a risk of contagion, not from whom or where. Methods such as differential privacy are used to ensure this. Cyber-resilience is high (+).

• No, data are (re)identifiable owing to the level of data collected, and stored centrally. Locations of contacts are also available. Cyber-resilience is low (–).

\item Can users \emph{erase the data}?

• Yes, they can do so at will; all data are deleted at the end point (+).

• No, there is no provision for data deletion, nor a guarantee that it can ever be deleted (–).

\item Is the \emph{purpose of data collection} defined?

• Yes, explicitly; for example, to alert users that they have encountered a potentially infected person (+).

• No, the purposes of data collection are not explicitly defined (–).

\item Is the purpose \emph{limited}?

• Yes, it is used for tracing and tracking of COVID-19 only (+).

• No, it can be regularly updated to add extra features that extend its functionality (–).

\item Is it \emph{used only for prevention}?

• Yes, it is used only to enable people voluntarily to limit spread (+).

• No, it is also used as a passport to enable people to claim benefits or return to work (–).

\item Is it used for \emph{compliance}?

• No, it is not used to enforce behaviour (+).

• Yes, non-compliance can result in punishment such as a fine or jail time (–).

\item Is it \emph{open-source}?

• Yes, the code is publicly available for inspection, sharing and collaborative improvement (+).

• No, the source code is proprietary, and no information about it is provided (–).

\item Is it \emph{equally available}?

• Yes, it is free and distributed to anyone (+).

• No, it is arbitrarily given only to some (–).

\item Is it \emph{equally accessible}?

• Yes, it is user-friendly, even for naive users, and works on the widest possible range of mobile phones (+).

• No, it can be used only by those with specific devices and with sufficient digital education (–).

\item Is there a \emph{decommissioning process}?

• Yes, there is a process for shutting it down (+).

• No, there are no policies in place (–).
\end{enumerate}
\end{framed}

%% file: priv+data.tex

\section{Privacy and Data Protection}

Privacy-related issues for \Covid mitigation strategies have triggered heated discussion, and at some point gained much of the media coverage.
This is understandable, since privacy and data protection is an important aspect of medical information flow, even in ordinary times: 

\MS{%
\begin{myquote} While public-health and safety amidst the health crisis is imperative, the data rights and privacy policy responses are critical, \emph{now, and after the crisis}. \cit{Transparency-key-2020-04-27}
\end{myquote}
}

\noindent
Moreover, the IT measures against \Covid are usually designed by computer scientists and specialists, for whom technological requirements (such as information security) are relatively easy to identify and understand.
As the authors of \cit{Brookings-Tech-Stream-CTAs-2020-04-27} cautiously stated:
\begin{myquote}
Some of the contact-tracing frameworks have been \emph{designed with security and privacy in mind}, to some degree.
\end{myquote}

\subsection{Privacy-Related Requirements}

The designers of contact-tracing apps have sometimes made overly bold statements about their privacy level. Nevertheless, those indicate a general requirement that is clearly important for many users:
\begin{myquote}
Any \emph{information} you choose to submit \emph{is protected at all times}.
\cit{NCSC-nhs-explainer-2020-05-04}
\end{myquote}

\noindent
In particular:

\begin{myquote}
developers of contact-tracing apps (…) should include recommendations for \emph{how back-end systems should be secured} and \emph{how long data should be retained}, criteria for \emph{what public health entities can qualify to use these technologies}, and explicit app store policies for \emph{what additional information}, such as GPS or government ID numbers, \emph{can be collected}. They should adopt commonly accepted practices such as security auditing, bug bounties, and abusability testing to \emph{identify vulnerabilities and unintended consequences} of a potentially global new technology. Finally, app creators—as well as the platforms that enable these applications—should make \emph{explicit commitments} for when these apps and their underlying APIs will be \emph{sunsetted}.
\cit{Brookings-Tech-Stream-CTAs-2020-04-27}
\end{myquote}

\begin{myquote}
The design is intended to mitigate \emph{privacy concerns inherent in a technological approach} to identifying possible new infections, the tech companies have said. They say the apps would not \emph{collect location data} or \emph{save records to a central database} and would instead rely on the records saved and distributed across people’s phones, which would be \emph{privacy-protected} and \emph{periodically erased}.
\cit{WashingtonPost-Most-Americans-2020-04-29}
\end{myquote}

\begin{myquote}
The app itself doesn’t collect information that would \emph{obviously reveal someone’s identity}
\cit{Wired-Just-how-2020-05-12}
\end{myquote}

\begin{myquote}
But under data protection laws the app isn’t \emph{anonymous}. GDPR and the UK’s data protection rules define ‘personal data’ as something that can identify an individual. Under GDPR, an identifier assigned to a phone can be considered personal data. (In the past a person’s IP address has been ruled to be personal data). While the Bluetooth logging system in the NHS app doesn’t collect location information, or other types of data, it does create an identifier (known as InstallationID) for every phone that uses the app. This counts as something that could \emph{lead to the identification of an individual}.
\cit{Wired-Just-how-2020-05-12}
\end{myquote}

\begin{myquote}
“The NHSX app does not preserve the \emph{anonymity of users}, as it primarily processes {pseudonymous}, not anonymous, personal data,” Michael Veale, a lecturer in digital rights and regulation at University College London, wrote in an analysis of the NHS app. “Anonymous information is only that which is not personal data”.
\cit{Wired-Just-how-2020-05-12}
\end{myquote}

\begin{myquote}
From the user perspective there is also the problem of \emph{informed consent}.
\cit{Guardian-democracies-2020-06-16}
\end{myquote}

\subsubsection{Risks and Threats}

Alerts can be \emph{too revealing}~\cit{BBC-Coronavirus-privacy-2020-03-05}, e.g.,
\begin{myquote}
it may be possible to work out \emph{who is associating with whom} from the app's ID numbers.
\cit{Register-UK-finds-2020-05-05}
\end{myquote}

\begin{myquote}
[the data from the app] may be \emph{linked to other datasets} at some point in future.
\cit{Gizmodo-UKs-Contact-Tracing-2020-05-13}
\end{myquote}

\noindent
A direct exploitation of personal information is even more dangerous:
\begin{myquote}
Kuwait and Bahrain have rolled out some of the \emph{most invasive} Covid-19 contact-tracing apps in the world, putting the \emph{privacy and security of their users at risk}, Amnesty International says.
\cit{BBCNews-Alarm-Kuwait-2020-06-16}
\end{myquote}

\begin{myquote}
The rights group found the apps were carrying out \emph{live or near-live tracking of users' locations} by uploading GPS co-ordinates to a central server.
(...)
The researchers say Bahraini and Kuwaiti authorities would easily be able to \emph{link this sensitive personal information to an individual}, as users are required to register with a national ID number.
\cit{BBCNews-Alarm-Kuwait-2020-06-16}
\end{myquote}

\begin{myquote}
Bahrain's app was linked to a television show called "Are You At Home?", which offered prizes to users who stayed at home during Ramadan.
\cit{BBCNews-Alarm-Kuwait-2020-06-16}
\end{myquote}

\MS{
\begin{myquote} [Australian government] also has a history of \emph{'mission creep'} regarding tracking mechanisms such as the metadata retention regime, where it allowed \emph{citizen’s data to be accessed by all sorts of law enforcement mechanisms}, contrary to initial intentions. \cit{Transparency-key-2020-04-27}
\end{myquote}
}

\subsubsection{Reasonable Privacy}

Most people accept that it might be necessary to waive users’ privacy in the short term in order to contain the epidemic. 
However, one must look for mechanisms that impact privacy to the least possible extent.
\begin{myquote}
The [Apple and Google] approach is not without risk, researchers acknowledge, but \emph{exploiting that risk would require significant effort by hackers for seemingly little reward}. For example, they would have to turn on a different phone every time they came near a different person and wait several days to see if it reported a positive test result.
\cit{Nature-Can-they-slow-2020-05-19}
\end{myquote}
\WJ{This closely resembles the notion of \emph{rational security} by Anderson and Moore~\cite{Anderson07incentives,Moore11economics}}

\noindent
Moreover, privacy-related requirements depend on the geographical and political context:
\begin{myquote}
The issues uncovered by Amnesty's investigation are particularly alarming given that the human rights records of Gulf governments are poor.
(...)
"If privacy is violated in a country like Norway, I can resort to regional tools such as the European Court of Human Rights and European Committee of Social Rights. But in our region there is not any such tool. On the contrary, resorting to local authorities may present an additional risk" [Mohammed al-Maskati, a Bahraini activist said.]
\cit{BBCNews-Alarm-Kuwait-2020-06-16}
\end{myquote}

\subsection{Data Protection and Potential Misuse of Data}

Here, the key questions are:
\begin{myquote}
\emph{What data} will they collect, and \emph{who is it shared with}?
\cit{MIT-covid-tracing-2020-05-07}
\end{myquote}

\noindent
\MS{%
as well as
%
\begin{myquote} \emph{how data is collected, stored and deleted}. \cit{Transparency-key-2020-04-27}
\end{myquote}
}

\noindent
In particular, it is often postulated to have
\begin{myquote}
less \emph{state access} and \emph{control over user data}
\cit{Telecoms-Unlike-France-2020-04-27}
\end{myquote}

\begin{myquote}
the \emph{limits on the type of data collection} are the core concern for states.
\cit{Politico-Google-Apple-2020-06-10}
\end{myquote}

\begin{myquote}
In Singapore, for example, the TraceTogether app can be used \emph{only by its health ministry} to access data. It assures citizens that the data is to be \emph{used strictly for disease control} and will not be \emph{shared with law enforcement agencies} for enforcing lockdowns and quarantine.
\cit{BBC-News-Why-Indias-2020-05-15}
\end{myquote}

\begin{myquote}
[In Australia,] \emph{Only health officials} in the states \emph{can access the data}, and you can’t be forced to download it.
\cit{Guardian-Covidsafe-2020-05-23}
\end{myquote}

\begin{myquote}
The app does not \emph{collect any of your personal data}.
\cit{NCSC-nhs-explainer-2020-05-04}
\end{myquote}

\subsubsection{Risks and Threats}

\begin{myquote}
the lack of \emph{clear privacy policies} and the use of centralised data storage increases the possibility that the \emph{data may be vulnerable to abuse}.
\cit{Top10VPN-Digital-Rights-2020-06-10}
\end{myquote}

\begin{myquote}
collection of data on centralised servers: Aside from the risk to privacy, \emph{collecting millions of datasets of personal information in a single place could be viewed as somewhat of a treasure trove}.
\cit{Telecoms-UK-2020-04-28}
\end{myquote}

\begin{myquote}
if the [central] database is hacked, the \emph{anonymity} provided by rotating pseudonyms is nullified, and \emph{individuals can be more easily tracked}. Plus, says Kreps, “there’s a risk of \emph{function creep} and \emph{state surveillance}”. “I have little faith in {government’s ability to keep data like this secure},” says Green.
\cit{Nature-Can-they-slow-2020-05-19}
\end{myquote}

\begin{myquote}
\emph{data breaches} can also come through \emph{cyberattacks} or \emph{independent actors within an agency}
\cit{Politico-Privacy-fears-2020-06-04}
\end{myquote}

\noindent
In particular, the collected information should not be  exploited for commercial purposes:
\begin{myquote}
existing regulations don’t address whether \emph{data can be shared across agencies} or if it can be \emph{sold by a third party} for non-Covid-19 tracking.
\cit{Politico-Privacy-fears-2020-06-04}
\end{myquote}

\begin{myquote}
We found code relating to \emph{Google’s advertising and tracking platforms} in 17 contact tracing apps. (…) Aside from the \emph{ethics of monetizing public health} in this way, the presence of such tracking code in contact tracing apps raises privacy red flags due to the \emph{targeting options} offered by Google’s ad platforms.

(...)
We also found code that enabled \emph{varying levels of integration with Facebook} in seven apps. This ranges from direct integration with Facebook’s advertising platform to functionality allowing users of the apps to link their Facebook accounts, or to share content from the contact tracing apps to Facebook.
\cit{Top10VPN-Digital-Rights-2020-06-10}
\end{myquote}

\noindent
\MS{A mistake in implementation can have serious privacy consequences in improperly tested software: 
\begin{myquote}
technical issues meant a person completing a form for the survey had been able to view information previously submitted by other users.
\cit{Bug-found-in-contact-tracing-2020-09-10}
\end{myquote}
}
\subsubsection{Impact on Society}

\begin{myquote}
\emph{Protections need to be put in place} to expressly \emph{prohibit economic and social discrimination} on the basis of information and technology designed to address the pandemic.
\cit{Brookings-Tech-Stream-CTAs-2020-04-27}
\end{myquote}

\begin{myquote}
{protecting} those communities who can be (…) harmed by the \emph{collection and exploitation of personal data}.
\cit{Brookings-Tech-Stream-CTAs-2020-04-27}
\end{myquote}

\subsection{Temporal Requirements, Safeguards and Limitations, Monitoring of Privacy and Data Abuse}

In this respect, an important question is:
\begin{myquote}
\emph{How will that information be used in the future?}
Are there \emph{policies in place to prevent abuse?}
\cit{MIT-covid-tracing-2020-05-07}
\end{myquote}

\begin{myquote}
To capture this information, guided by principles put forward by the American Civil Liberties Union and others, we asked five questions: Is it \emph{voluntary}? Are there limitations on \emph{how the data gets used}? Will data be \emph{destroyed} after a period of time? Is data collection \emph{minimized}? Is the effort \emph{transparent}?
\cit{MIT-covid-tracing-2020-05-07}
\end{myquote}

\noindent
Among temporal requirements, sunsetting is most often mentioned.
\begin{myquote}
There is also a concern that the [surveillance] technology \emph{will continue to be used after the threat of the coronavirus recedes}
\cit{BBCNews-Alarm-Kuwait-2020-06-16}
\end{myquote}

\begin{myquote}
questions about privacy — including \emph{when the tracking will be switched off}.
\cit{POLITICO-Polands-2020-04-02}
\end{myquote}

\begin{myquote}
"We need \emph{an independent state figure that's not the government who can guarantee this data will eventually be deleted}," said Arnoldo Frigessi, a professor at the University of Oslo (…)"We need to ask the questions: \emph{When will this stop, and who will get to decide?}"
\cit{POLITICO-Polands-2020-04-02}
\end{myquote}

\begin{myquote}
Any information you submit is \emph{deleted once it is no longer needed to help manage the spread of coronavirus}.
\cit{NCSC-nhs-explainer-2020-05-04}
\end{myquote}

\begin{myquote}
All the data will stay with the government for six years”
\cit{POLITICO-Polands-2020-04-02}
\end{myquote}

\noindent
For example, the Norwegian app Smittestopp explicitly promised the following sunsetting requirements~\cit{helsenorge-Together-2020-04-28}:
\begin{itemize2}
\item data from the last 30 days are constantly recorded and \emph{older data are deleted}
\item one can \emph{delete one's personal information} at any time
\item one can \emph{delete the app}.
\item the user can also \emph{choose whether to turn the logging features on or off}
\item one has the \emph{right to access the data} that the Norwegian Institute of Public Health stores about them.
\end{itemize2}

\noindent
Provisions for monitoring of privacy are also essential:
\begin{myquote}
[Unlinkability] must be backed up with clear lines of \emph{accountability}, processes for \emph{evaluating linkage or export requests}, and \emph{strong assurance monitoring}.
\cit{Gizmodo-UKs-Contact-Tracing-2020-05-13}
\end{myquote}

\begin{myquote}
\emph{judicial oversight} and \emph{sunset provisions} (…) safeguards with respect to the privacy of data
\cit{Brookings-Tech-Stream-CTAs-2020-04-27}
\end{myquote}

\subsection{Impact of Privacy on Health Protection}

\begin{myquote}
As the virus spreads rapidly, it's vital that the public \emph{are giving the information [people] need to protect themselves and others}.
\cit{BBC-Coronavirus-privacy-2020-03-05}
\end{myquote}

\begin{myquote}
Dr Lee says the public needs to remain mature with this information - otherwise "people who fear being judged will hide and this will \emph{put everyone in further danger}".
\cit{BBC-Coronavirus-privacy-2020-03-05}
\end{myquote}

\subsection{Impact on Psychological Wellbeing and Society}

\begin{myquote}
the government is letting people know if they were in the vicinity of a patient (…) and now there is as much \emph{fear of social stigma} as of illness
\cit{BBC-Coronavirus-privacy-2020-03-05}
\end{myquote}

\begin{myquote}
even if patients are not outright \emph{identified}, they're \emph{facing judgement - or ridicule} - online.
\cit{BBC-Coronavirus-privacy-2020-03-05}
\end{myquote}

\begin{myquote}
A research team at Seoul National University's Graduate School of Public Health (…) found \emph{"criticisms and further damage" were more feared than having the virus}.
\cit{BBC-Coronavirus-privacy-2020-03-05}
\end{myquote}

\begin{myquote}
Doctors warn that online pursuit of patients could have very serious consequences. \emph{Malicious comments online have long been a problem in South Korea}, and in some cases have led to \emph{suicide}.
\cit{BBC-Coronavirus-privacy-2020-03-05}
\end{myquote}

\subsection{Privacy vs.~Epidemiological Efficiency}
\label{sec:priv-vs-epid}

Much of the debate of centralized vs.~decentralized design for contact-tracing apps follows the dilemma whether it is justified to sacrifice the users' privacy for the sake of efficiency in containing the epidemic:
\begin{myquote}
[A centralized contact-tracing app] takes data from people's phones and saves it on a central system where experts are trusted to \emph{make the best possible use of the data}, including providing \emph{advice to people} as and when necessary.
\cit{Register-UK-finds-2020-05-05}
\end{myquote}

\begin{myquote}
Some [centralized apps] store all users’ interaction data on government servers that \emph{analyse the data} and \emph{perform the contact matching}. Proponents say that this ‘centralized’ model allows health authorities to use the database to \emph{piece together a view of the network of contacts}, enabling further \emph{epidemiological insights} such as \emph{revealing clusters} and \emph{superspreaders}.
\cit{Nature-Can-they-slow-2020-05-19}
\end{myquote}

\begin{myquote}
[decentralized contact-tracing apps:] puts users in \emph{more control of their information}, and alerts them automatically with \emph{no intervention from a third party}.
\cit{Register-UK-finds-2020-05-05}
\end{myquote}

\begin{myquote}
while the (…) decentralized model \emph{protects people's privacy}, it leaves the authorities blind. It puts a public health disaster outside the reach of those who can \emph{help most thorough analysis of the population}.
\cit{Register-UK-finds-2020-05-05}
\end{myquote}

\begin{myquote}
Health experts are less in favour of [the decentralized] model because it makes it harder to \emph{spot outbreaks} and is harder to \emph{follow up for close contacts}, but privacy advocates insist it is the most \emph{secure}.
\cit{Guardian-Covidsafe-2020-05-23}
\end{myquote}

\subsection{Privacy {in Support of} Epidemiological Efficiency}
\label{sec:priv-support-epid}

As we have seen in Section~\ref{sec:priv-vs-epid}, there is a tradeoff between protecting privacy vs.~collecting and protecting all the information that can be useful in fighting the epidemic.
The relationship is not that simple, though.
Privacy provisions are instrumental in building trust. Conversely, lack of privacy undermines trust, and may weaken the epidemiological, economic, and social effects of the mitigation activities.

\begin{myquote}
Privacy fears \emph{threaten} New York City's \emph{coronavirus tracing efforts}
\cit{Politico-Privacy-fears-2020-06-04}
\end{myquote}

\begin{myquote}
the de Blasio administration’s \emph{unwillingness to specify} how privacy will be protected will \emph{limit the tracing effort's reach} and potentially \emph{prolong the need for strict lockdown}
\cit{Politico-Privacy-fears-2020-06-04}
\end{myquote}

\begin{myquote}
[Manual] Contact tracing relies on a city worker building \emph{trust} with a Covid-19-positive person, who then details where they’ve been and with whom they’ve been in contact prior to the onset of symptoms
(...)
Contact tracing requires handing over intimate personal data — including home addresses, names of friends and relations — to strangers, many of whom were only recently trained and hired to collect the information.
\cit{Politico-Privacy-fears-2020-06-04}
\end{myquote}

\begin{myquote}
Hundreds of thousands of New Yorkers will be asked to disclose personal information this month as part of the city’s herculean Covid-19 tracing effort — but \emph{suspicions over how the government will use that information} are threatening the city’s best chance to \emph{crawl out of its coronavirus lockdown}.
\cit{Politico-Privacy-fears-2020-06-04}
\end{myquote}

\begin{myquote}
“I did already hear rumors in Yiddish that they shouldn’t answer the [contact tracing] calls,” said Yosef Hershkop, manager of Kāmin Health Crown Heights Urgent Care. “There’s no way they’re going to \emph{get} any more \emph{cooperation} from the Jewish community than the African American community or Arab community or Muslim community. Everyone is \emph{suspicious}.”
\cit{Politico-Privacy-fears-2020-06-04}
\end{myquote}

%% file: user+tech.tex

\section{User-Related Aspects}

The measures must be adopted and followed by the people, in order to make them effective.

\subsection{User Incentives}
\label{sec:userinc}

\MS{%
\begin{myquote}
The design, implementation and adoption of complex socio-technical systems must be \emph{driven by the interests of users}. \cit{Transparency-key-2020-04-27}
\end{myquote}
}

\begin{myquote}
Ultimately, contact tracing 
(...)
can reduce the spread of disease through the population, but does not confer direct protection on any individual. This creates \emph{incentive problems} that need careful thought: \emph{What is in it for the user} who will sometimes be instructed to miss work and avoid socializing, but does not derive \emph{immediate benefits} from the system?
\cit{Brookings-Tech-Stream-CTAs-2020-04-27}
\end{myquote}

\begin{myquote}
A poll published last week by the Kaiser Family Foundation found the nation similarly split on \emph{willingness to use} an infection-alert app but found much \emph{higher rates of acceptance} — with 2 out of every 3 Americans willing — when they were told the technology was seen as \emph{assisting in opening schools and businesses} and \emph{helping revive the stalled national economy}. But fewer than 3 in 10 were willing to use an app if there was a “chance the data could be hacked.”
\cit{WashingtonPost-Most-Americans-2020-04-29}
\end{myquote}

\begin{myquote}
The government (…) didn't exclude mandatory downloading [that] stands in stark contrast to the 'smart lockdown' with \emph{strong reliance on personal responsibility}
\cit{euobserver-Dutch-soap-opera-2020-05-07}
\end{myquote}

\begin{myquote}
Won't this tool \emph{divert attention from more important measures}, and \emph{make people less alert}?
\cit{Panoptykon-ProteGo-Safe-2020-05-06}
\end{myquote}

\begin{myquote}
even with a very accurate test, the fewer people in a population who have a condition, the more likely it is that an individual's positive result is wrong. If it is, people might think they have the antibodies (and thus may have immunity), when in fact they do not.
\cit{ScientificAmerican-AntibodyTests-2020-06-17}
\end{myquote}

\subsection{User Experience}
\label{sec:userx}

\begin{myquote}
Since the app was released in mid-March \emph{people's experiences have varied}, with many willing to give up their personal information to help reduce the number of overall infections. Most said the app is \emph{prone to failure}, not a surprise as the Polish government created it in just three days based on an out-of-the-box service offered by a third-party developer.
\cit{POLITICO-Polands-2020-04-02}
\end{myquote}

\noindent
\MS{
\emph{Interoperability} is an important aspect of user experience, so that:
\begin{myquote}
Users of contact tracing and warning apps can now \emph{use a single app} when they travel cross-borders and still benefit from contact tracing. \cit{EU-Commission-implements-2020-10-28}
\end{myquote}
}
\MS{%
\begin{myquote}
This means a user of one of those apps who travels to any of the other countries can expect their national app to send relevant exposure notifications in the same way it should if they had not travelled — \emph{without the need to download any additional software}. \cit{EU-switches-on-2020-10-19}
\end{myquote}
}

\subsection{Adoption and Its Impact on Other Requirements}

The aspects indicated in Sections~\ref{sec:userinc} and~\ref{sec:userx} have impact on adoption of the strategy:
\begin{myquote}
How many people will \emph{download and use} [the apps], and how widely used do they have to be in order to \emph{succeed}?
\cit{MIT-covid-tracing-2020-05-07}
\end{myquote}

\begin{myquote}
Another challenge is ensuring that \emph{enough people download the app to make it effective}.
\cit{Nature-Can-they-slow-2020-05-19}
\end{myquote}

\noindent
For example, it was reported in~\cit{LeMonde-StopCovid-2020-06-10} that only 2\% of people in France installed the app as of 10/06.
\WJ{Mention the uptake in Germany!}
This has important consequences:
\begin{myquote}
Reduced adoption [of the app] could limit its \emph{effectiveness in slowing new infections and deaths}.
\cit{WashingtonPost-Most-Americans-2020-04-29}
\end{myquote}


\begin{myquote}
The Oxford models\footnote{ \cite{Hinch20covid-uptake} }
found that "the app \emph{has an effect} at all \emph{levels of uptake}"
(...)
The paper says that if 80\% of all smartphone users download the app—a number that excludes groups less likely to have a smartphone and is equivalent to 56\% of the overall population—this would be enough to \emph{suppress the pandemic} on its own, without any other form of intervention.
\cit{MIT-Technology-coronavirus-apps-2020-06-05}
\end{myquote}

\begin{myquote}
if fewer people download the app, say the researchers, \emph{other prevention and containment measures will be required}. These include social distancing, widespread testing, manual contact tracing, medical treatment, and regional shutdowns
\cit{MIT-Technology-coronavirus-apps-2020-06-05}
\end{myquote}

\noindent
Trust is an important issue (see also the connection between privacy provisions and trust in Section~\ref{sec:priv-support-epid}).
\begin{myquote}
For people to use the app, there \emph{needs to be trust}.
\cit{Wired-Just-how-2020-05-12}
\end{myquote}

\begin{myquote}
[J.T. Lane, chief population officer at the Association of State and Territorial Health Officials:] “You need to learn everything you can about [the users’] culture and maybe even their previous relationship with technology,” Lane said, including “how stigma and discrimination might be playing a role in that particular group or culture.”
\cit{POLITICO-States-struggle-2020-05-17}
\end{myquote}

\MS{%
\begin{myquote}
[Approach of] blending technology with manual contact tracing, solves two key problems with contact tracing apps: trust and efficacy. \emph{Trust is a key foundation of public health} and a big problem for contact tracing apps; (...) research shows that people are divided on \emph{whom they trust to provide the apps} as well as whether \emph{they trust the apps to actually work} and \emph{to protect their data}. \cit{How-to-Fix-COVID-2020-12-07}
\end{myquote}
}

\section{Technological Requirements}

The concrete measures have to \emph{work}, in the most basic sense of the term.

\subsection{Basic Requirements}

First of all, this means that the design correctly addresses the desired functionalities, the implementation is correct with respect to the design, and the measures have an acceptable level of usability.

\begin{myquote}
UK finds itself almost alone with centralized virus contact-tracing app that \emph{probably won't work well}, asks for your location, may be illegal
\cit{Register-UK-finds-2020-05-05}
\end{myquote}

\begin{myquote}
According to local reports from folks that have downloaded the program thus far, not only is the app \emph{incompatible with a range of different devices}, but it’s also a \emph{battery-draining} mess that can be \emph{confusing to use}.
\cit{Gizmodo-UKs-Contact-Tracing-2020-05-13}
\end{myquote}

\begin{myquote}
Beyond concerns over privacy, one key practical challenge to phone-based contact tracing is \emph{making accurate measurements} of how close two devices are. \cit{Nature-Can-they-slow-2020-05-19}
\end{myquote}

\subsection{Transparency}

Moreover, the design and the implementation should be sufficiently transparent. 

\MS{%
\begin{myquote}
The forthcoming challenge for the Australian government is to be \emph{open} and \emph{transparent} about the app’s privacy implications. (...) the public deserved to be \emph{fully informed about the cyber security implications}. \cit{Transparency-key-2020-04-27}
\end{myquote}
}

\MS{%
\begin{myquote}
It’s essential to see \emph{transparency} around \emph{encryption methods used by the app, the potential blue tooth (sic) vulnerabilities related to the collection of data, and how access to the data by state and territory health officers will be audited}. \cit{Transparency-key-2020-04-27}
\end{myquote}
}

\noindent
Covid Tracing Tracker puts forward a number of ``basic questions'' related to the transparency of the design and implementation:
\begin{myquote}
Who is producing [the app]? Is it released yet? Where will it be available, and on what platforms? What technologies does it use? (…) how many people have downloaded it and what level of penetration it has achieved.
How \emph{transparent} are the makers about their work?
\cit{MIT-covid-tracing-2020-05-07}
\end{myquote}

\noindent
\MS{%
Other important aspects that help to build trust in digital solutions are: \emph{transparency of the code} and the possibility to \emph{verify the code} by the public and experts \cit{SDZ-Corona-App-2020-05-06}, see also this quote on the Australian Government's COVIDSafe application:
\begin{myquote}
The code is \emph{not publicly available} and open-source data is \emph{stored on private, cloud-based web servers}, which provides a \emph{honey-pot for cyber-attacks} on this highly sensitive information about the population. \cit{Transparency-key-2020-04-27}
\end{myquote}
}

\subsection{Interoperability}

\MS{An important technical goal that demands international cooperation in early design phases is \emph{cross-border interoperability}.}

\MS{%
\begin{myquote}
Many [EU] member states have implemented \emph{national} contact tracing and warning applications. It is now time to make them \emph{interact with each other}. \cit{EU-tests-platforms-2020-04-14}
\end{myquote}
}

\noindent
The goals include relaxation of the economic and social restrictions, as well as coordinated data protection:
\MS{%
\begin{myquote}
In time, this will \emph{simplify cross-border contact tracing} and \emph{allow our people to visit our neighbours for work, trade or leisure}. \cit{Cybernetica-Proposes-2020-05-06}
\end{myquote}

\begin{myquote}
This \emph{coordinated approach} will also help to ensure that \emph{the same level of data protection is given to users [across EU]}. \cit{EU-Commission-implements-2020-10-28}
\end{myquote}
}

\noindent
Interoperability-related requirements concern data flow, efficiency, and privacy:
\MS{%
\begin{myquote}
Each [national] {backend} communicates only with the gateway, so you don’t have a direct exchange between the different backends. 
(...) And that \emph{keeps the data flow to a minimum} and is much more \emph{efficient}. \cit{The-race-is-on-2020-10-19}
\end{myquote}
}

\MS{%
\begin{myquote}
\emph{No other information} than arbitrary keys, generated by the apps, will be \emph{handled by the gateway}. (...) The information is \emph{pseudonymised, encrypted, kept to the minimum and only stored as long as necessary} to trace back infections. It \emph{does not allow the identification of individual persons}, nor to \emph{track location or movement} of devices. \cit{EU-switches-on-2020-10-19}
\end{myquote}
}

\noindent
Risks and threats:
\MS{%
\begin{myquote}
Ensuring {interoperability} is \emph{technically difficult} and may  \emph{requirechanges to existing national apps}. \cit{EU-Commission-implements-2020-10-28}
\end{myquote}
}

\MS{%
\begin{myquote}
Run from the European Commission’s data centre in Luxembourg, the system was designed by SAP and T-Systems, using a (...) \emph{third-party server} dubbed a “{gateway}” to let [EU] member countries’ apps share data. \cit{The-race-is-on-2020-10-19}
\end{myquote}
}

\noindent 
\MS{As with the national contact tracing app and user adoption
\begin{myquote}
The effectiveness of this interoperable gateway will depend on how many [EU] Member States register their national contact tracing apps with this new service. \cit{EU-Commission-implements-2020-10-28}
\end{myquote}
}

%% file: eval+know.tex

\section{Evaluation and Learning for the Future}

\Covid mitigation activities should be subject to rigorous assessment. Moreover, their outcomes should be used to extend our knowledge about this and similar pandemics, and better defend ourselves in the future.

\subsection{Evaluation-Related Requirements}

Systematic evaluation of the adopted solutions is itself a requirement:
\begin{myquote}
''After the spread of virus ends," [Mr Goh, from the Korea Centers for Disease Control Prevention] says, "there has to be \emph{society's assessment whether or not this was effective and appropriate}."
\cit{BBC-Coronavirus-privacy-2020-03-05}
\end{myquote}

\begin{myquote}
A \emph{review} and \emph{exit strategy} must be in place to establish when and how fast this should happen. These \emph{assessments} should be \emph{conducted by an independent body}, such as a regulator or an ethics advisory board, and \emph{not by the designers or the government itself}. Circumstances and attitudes are changing quickly, so the questions (...) must be \emph{asked anew at regular intervals}.
\cit{Nature-Comment-Ethical-guidelines-2020-06-04}
\end{myquote}

\noindent
Evaluation and assessment is bound to be multi-dimensional and complex:
\begin{myquote}
Before rolling out any app, the EU needs to do a \emph{coordinated assessment} of \emph{usefulness, effectiveness, technological readiness, cyber security risks and threats to fundamental freedoms and human rights}.
\cit{euobserver-Dutch-soap-opera-2020-05-07}
\end{myquote}

\begin{myquote}
it's difficult to \emph{determine the effect of app use in combination with} early and extensive testing, the massive wearing of face masks and strict enforcement at physical distance.
\cit{euobserver-Dutch-soap-opera-2020-05-07}
\end{myquote}

\subsection{Learning for the Future}

Last but not least, we need to building scientific knowledge, and learn how to better fight pandemics in the future.

\begin{myquote}
Anonymised data on movement patterns of the population collected by the app are also supposed to be used to \emph{develop efficient infection control measures}. [They] will be used to \emph{gain insight into the effect of changes to the measures for fighting the virus}.
\cit{helsenorge-Together-2020-04-28}
\end{myquote}